\documentclass[preprintnumbers,amsmath,amssymbm,prd]{revtex4}
\usepackage{epsfig}
\usepackage{graphicx}
\usepackage{amssymb}

\begin{document}
\title{The superradiant instability spectrum of the hydrodynamic vortex model}
\author{Shahar Hod}
\affiliation{The Ruppin Academic Center, Emeq Hefer 40250, Israel}
\affiliation{ } \affiliation{The Hadassah Institute, Jerusalem
91010, Israel}
\date{\today}

\begin{abstract}
\ \ \ We study analytically the superradiant instability properties of
the hydrodynamic vortex model, an asymptotically flat acoustic
geometry which, like the spinning Kerr black-hole spacetime,
possesses an effective ergoregion. In particular, we derive a
compact analytical formula for the complex resonant frequencies that 
characterize the long-wavelength dynamics of sound modes in
this physically interesting acoustic spacetime. 
\end{abstract}
\bigskip
\maketitle


\section{Introduction}

The canonical Kerr black-hole solution of the vacuum Einstein field equations \cite{Kerr} 
is known to possess an {\it ergoregion} \cite{Chan}, a
spacetime region which extends from the horizon
$r_+=M+\sqrt{M^2-a^2}$ to
$r_{\text{ergo}}=M+\sqrt{M^2-a^2\cos^2\theta}$, in 
which all physical observers are inevitably dragged by the rotation
of the black hole [Here $\{M,a\}$ are respectively the mass and
angular-momentum per unit mass of the Kerr black hole and $\theta$
is the polar angle of the stationary axisymmetric spacetime].

As pointed out by Zel'dovich \cite{Zel} and by
Press and Teukolsky \cite{PressTeu1,PressTeu2}, a co-rotating
bosonic wave field that propagates in the black-hole ergosphere can
be superradiantly amplified (that is, can extract rotational energy
from the central spinning black hole) if its proper frequency is bounded from above
by the relation $\omega<m\Omega_{\text{H}}$, where $m$ is the
azimuthal harmonic index of the bosonic field mode and
$\Omega_{\text{H}}$ is the angular velocity of the black-hole horizon \cite{Chan}.

It is interesting to emphasize the fact that, although bosonic fields
can be superradiantly amplified in the ergoregion, the asymptotically flat Kerr
black-hole spacetime is known to be stable under perturbations of
massless bosonic fields \cite{PressTeu2,Whit}. This important physical 
property of the composed Kerr-black-hole-massless-bosonic-field 
system is closely related to the well known absorption
properties (ingoing boundary conditions) that characterize the classical black-hole horizon. 
In particular, the central black hole swallows (and also scatters away to infinity) the potentially
dangerous amplified bosonic fields before they have the chance to
develop exponentially growing instabilities inside the ergosphere. 

It is worth mentioning that, in order to trigger superradiant instabilities
in the spacetime of a spinning Kerr black hole, some additional
confinement mechanism (which can be provided, for example, by a
reflecting mirror which is placed around the central black hole \cite{PressTeu1,Ins1} or, as in the 
case of amplified massive bosonic fields, by the mutual gravitational attraction between the central
black hole and the fields \cite{Ins2}) is required in order to
prevent the superradiantly amplified bosonic fields from radiating
their energies to infinity \cite{Ins1,Ins2}.

Intriguingly, Friedman \cite{Fri} has pointed out the physically important fact that, 
as opposed to the Kerr black-hole spacetime, 
spinning horizonless spacetimes (and, in general, rotating physical
systems that have no absorptive boundaries) that possess
ergoregions in which bosonic fields can be superradiantly amplified 
may generally be unstable to co-rotating bosonic perturbation fields. 

In order to demonstrate this interesting physical phenomenon in the analogous setup of fluid dynamics \cite{Unr,Noteun}, 
the physical properties of the {\it hydrodynamic vortex} model have recently been studied numerically 
in the physically important work \cite{Cars}. 
This composed physical system describes a two-dimensional purely circulating flow of a vorticity free ideal
fluid that can be described by a non-trivial effective spacetime metric [see
Eq. (\ref{Eq3}) below]. 
In particular, this rotating acoustic spacetime has
no absorptive horizons but, like the familiar spinning Kerr
spacetime, it is characterized by the presence of an effective
acoustic ergoregion whose radial boundary is defined by the circle at which
the tangential velocity of the fluid equals the speed of propagating
sound waves in the fluid \cite{Sla} [see Eq. (\ref{Eq4}) below].

The highly interesting {\it numerical} results presented in \cite{Cars} for
the physical properties of the hydrodynamic vortex model have
established the fact that, in accord with the prediction of \cite{Fri}, 
this horizonless physical system may develop exponentially growing instabilities due to the 
superradiant amplification phenomenon of linearized sound waves in
the ergoregion of the effective spinning acoustic spacetime. 

The main goal of
the present paper is to explore, using {\it analytical} techniques, the superradiant instability
spectrum that characterizes the physically interesting hydrodynamic vortex model. 
In particular, as we shall explicitly prove below, the complex resonant frequencies that characterize the
dynamics of linearized sound waves in the spinning acoustic spacetime 
can be determined analytically in the dimensionless regime $C\omega\ll1$ of
small field frequencies, where $C$ is the radius of the effective acoustic ergoregion.

\section{Description of the system}

We shall analyze the dynamics of linearized sound waves in a
vorticity free barotropic ideal fluid. The background (unperturbed)
fluid velocity of a two-dimensional purely circulating flow is
characterized by the relations \cite{Cars}
\begin{equation}\label{Eq1}
v_r=v_z=0\ \ \ ; \ \ \ v_{\phi}=v_{\phi}(r)\  ,
\end{equation}
where $\{r,\phi\}$ are the radial and azimuthal coordinates in the
plane of flow, and $z$ denotes the third spatial coordinate which is
perpendicular to the plane (the $xy$ plane) of flow. A locally
irrotational (vorticity free) fluid flow is characterized by the
simple functional relation \cite{Cars}
\begin{equation}\label{Eq2}
v_{\phi}={C\over r}\
\end{equation}
for the tangential component of the velocity field, where the
proportionality constant $C$ characterizes the strength of
circulation in the fluid flow. The angular momentum conservation law
yields the relation \cite{Cars} $\rho v_{\phi}r=\text{constant}$,
which implies that the background density $\rho$ of the fluid is a
constant [see Eq. (\ref{Eq2})]. The assumption of a barotropic fluid
system then implies that the speed $c$ of linearized sound waves in the fluid and 
the background pressure $P$ of the fluid are also constants.

The circulating flow of the fluid in the hydrodynamic
vortex model produces an acoustic spacetime whose effective two-dimensional 
geometry is described by the line-element \cite{Cars,Sla,Fed,Dol}
\begin{equation}\label{Eq3}
ds^2=-c^2\Big(1-{{C^2}\over{c^2r^2}}\Big)dt^2+dr^2-2Cdtd\phi+r^2d\phi^2+dz^2\
.
\end{equation}
Interestingly, this rotating acoustic geometry is characterized by
the presence of an effective ergoregion whose radius
\cite{Cars,Sla,Fed,Dol}
\begin{equation}\label{Eq4}
r_{\text{ergo}}={{|C|}\over{c}}\
\end{equation}
is determined by the circle at which the tangential velocity of the
fluid [see Eq. (\ref{Eq2})] equals the speed $c$ of propagating
sound waves in the fluid. We shall henceforth use natural units in
which $c=1$ \cite{Notecc}.

As shown in \cite{Unr,Cars,Carn}, the dynamics of linearized
perturbation fields (sound modes) in the effective acoustic
spacetime (\ref{Eq3}) is mathematically governed by the familiar Klein-Gordon wave
equation \cite{Notefl}
\begin{equation}\label{Eq5}
\nabla^\nu\nabla_{\nu}\Psi={{1}\over{\sqrt{|g|}}}\partial_{\mu}\Big(\sqrt{|g|}
g^{\mu\nu}\partial_{\nu}\Psi\Big)=0\  ,
\end{equation}
where $g$ is the determinant of the effective line element
(\ref{Eq3}). Substituting the field decomposition \cite{Notecy}
\begin{equation}\label{Eq6}
\Psi(t,r,\phi,z)={{1}\over{\sqrt{r}}}\sum_{m=-\infty}^{\infty}\psi_m(r;\omega)e^{im\phi}e^{-i\omega
t}\
\end{equation}
into the Klein-Gordon wave equation (\ref{Eq5}) and using the line
element (\ref{Eq3}) of the effective two-dimensional acoustic spacetime, one finds
that the radial acoustic eigenfunctions $\psi_m(r;\omega)$ are determined by
the ordinary differential equation \cite{Notemp,Noteinv}
\begin{equation}\label{Eq7}
\Big[{{d^2}\over{dr^2}}+\Big(\omega-{{Cm}\over{r^2}}\Big)^2-{{m^2-{1\over
4}}\over{r^2}}\Big]\psi_m(r;\omega)=0\  .
\end{equation}

\section{Boundary conditions}

Taking cognizance of the relation (\ref{Eq2}) for the tangential
velocity field of the background fluid, one immediately realizes
that the hydrodynamic description breaks down on the symmetry axis
$r=0$ of the spacetime. In order to describe a physically realistic system, 
it has been suggested in \cite{Cars} to place an infinitely long supporting cylinder of
finite proper radius $R_0$ at the center of the dynamical fluid
system. In particular, as discussed in \cite{Cars,Carn}, the
physically motivated boundary condition for the effective scalar eigenfunction
$\Psi$ at the surface of the central supporting cylinder is given by the functional relation \cite{Cars,Carn}
\begin{equation}\label{Eq8}
{{{{d\Psi}/{dr}}}\over{\Psi}}(r=R_0)=-{{i\rho\omega}\over{Z_{\omega}}}\
,
\end{equation}
where $Z_{\omega}$ is the frequency-dependent impedance of the
cylinder \cite{Lax}, a physical parameter that quantifies the
interaction of the propagating sound wave with the material of the
scattering cylinder \cite{Cars,Carn,Lax}.

In addition, an asymptotically flat acoustic geometry is
characterized by the physical boundary condition of purely outgoing
waves at asymptotic infinity [see Eq. (\ref{Eq6})]:
\begin{equation}\label{Eq9}
\Psi(r\to\infty)\sim {{e^{i\omega r}}\over{\sqrt{r}}}\  .
\end{equation}
That is, we consider purely outgoing waves at large distances from the central cylinder.

Interestingly, the Schr\"odinger-like ordinary differential equation (\ref{Eq7}),
supplemented by the physically motivated boundary conditions
(\ref{Eq8}) and (\ref{Eq9}), determine the complex resonant frequencies $\{\omega(C,R_0,m)\}$
which characterize the dynamics of linearized sound waves in the
effective acoustic spacetime (\ref{Eq3}). It is worth emphasizing
that resonant field frequencies with $\Im\omega>0$ [see Eq.
(\ref{Eq6})] are associated with superradiantly {\it unstable} modes that 
grow exponentially in time. 
As we shall explicitly show in the next section, the complex resonant frequencies of the hydrodynamic
vortex model can be studied analytically in the dimensionless regime $C\omega\ll1$
of small field frequencies.

\section{The resonance equation and its regime of validity}

In the present section we shall analyze the Schr\"odinger-like
differential equation (\ref{Eq7}) which determines the radial
behavior of the acoustic eigenfunctions $\psi_m(r)$. As we shall explicitly prove below, 
the characteristic radial equation (\ref{Eq7}) can be
solved {\it analytically} in the two asymptotic radial regions $r\ll
m/\omega$ and $r\gg C$. We shall then show that, for small resonant
frequencies in the regime
\begin{equation}\label{Eq10}
C\omega\ll1\  ,
\end{equation}
one can use a functional matching procedure in the overlapping
region $C\ll r\ll m/\omega$ in order to determine analytically the complex
resonance spectrum $\{\omega(C,R_0,m;n)\}$ that characterizes the
dynamics of linearized sound waves in the hydrodynamic vortex model.

We shall first solve the Schr\"odinger-like differential equation
(\ref{Eq7}) in the radial region
\begin{equation}\label{Eq11}
r\ll m/\omega\  ,
\end{equation}
in which case one may approximate (\ref{Eq7}) by 
\begin{equation}\label{Eq12}
\Big[{{d^2}\over{dr^2}}+\Big({{Cm}\over{r^2}}\Big)^2-{{m^2-{1\over
4}}\over{r^2}}\Big]\psi_m=0\  .
\end{equation}
Here we have used the strong inequality
$\omega^2\ll m^2/r^2$ [see Eq. (\ref{Eq11})]. In addition, we have
used the strong inequality $Cm\omega/r^2\ll m^2/r^2$ which stems
from the small-frequency assumption $C\omega\ll 1\leq m$ [see Eq. (\ref{Eq10})].

The general mathematical solution of (\ref{Eq12}) can be expressed in terms of the Bessel
functions of the first and second kinds (see Eq. 9.1.53 of
\cite{Abram}):
\begin{equation}\label{Eq13}
\psi_m(r)=A_1r^{1\over 2}\cdot J_m\Big({{Cm}\over{r}}\Big)+A_2r^{1\over
2}\cdot Y_{m}\Big({{Cm}\over{r}}\Big)\ ,
\end{equation}
where the normalization constants $\{A_1,A_2\}$ are determined by
the physical boundary condition (\ref{Eq8}) of the wave field at the surface
$r=R_0$ of the central supporting cylinder. In particular, substituting
(\ref{Eq13}) into (\ref{Eq8}), one finds the dimensionless ratio [see Eq. (\ref{Eq6})]
\begin{equation}\label{Eq14}
{{A_2}\over{A_1}}=-{{J^{'}_m\big({{Cm}\over{R_0}}\big)-{{i\rho\omega
R^2_0}\over{Z_{\omega}Cm}}\cdot J_m\big({{Cm}\over{R_0}}\big)}\over{{Y^{'}_m\big({{Cm}\over{R_0}}\big)-{{i\rho\omega
R^2_0}\over{Z_{\omega}Cm}}\cdot Y_m\big({{Cm}\over{R_0}}\big)}}}\  ,
\end{equation}
where a prime $'$ denotes a derivative of the Bessel function with
respect to its argument $Cm/r$. Using Eq. 9.1.27c of \cite{Abram},
one can express the dimensionless ratio (\ref{Eq14}) in the form
\begin{equation}\label{Eq15}
{{A_2}\over{A_1}}=-{{{{R_0}\over{C}}\big(1+{{i\rho\omega
R_0}\over{Z_{\omega}m}}\big)\cdot J_m\big({{Cm}\over{R_0}}\big)-
J_{m-1}\big({{Cm}\over{R_0}}\big)}\over
{{{R_0}\over{C}}\big(1+{{i\rho\omega
R_0}\over{Z_{\omega}m}}\big)\cdot Y_m\big({{Cm}\over{R_0}}\big)-
Y_{m-1}\big({{Cm}\over{R_0}}\big)}}\  .
\end{equation}
Using the small-argument,
\begin{equation}\label{Eq16}
{{Cm}\over{r}}\ll1\  ,
\end{equation}
asymptotic behaviors of the Bessel
functions (see Eqs. 9.1.7 and 9.1.9 of \cite{Abram}), one finds from (\ref{Eq13}) the expression
\begin{eqnarray}\label{Eq17}
\psi_m(r)=A_1(m!)^{-1}\Big({{Cm}\over{2}}\Big)^mr^{-m+{1\over2}}
-A_2{\pi}^{-1}(m-1)!\Big({{Cm}\over{2}}\Big)^{-m}r^{m+{1\over2}}\
\end{eqnarray}
for the radial acoustic eigenfunction that characterizes the linearized
perturbation modes of the hydrodynamic vortex model in the
intermediate radial region [see Eqs. (\ref{Eq11}) and (\ref{Eq16})]
\begin{equation}\label{Eq18}
Cm\ll r \ll m/\omega\  .
\end{equation}

We shall next solve the Schr\"odinger-like radial differential equation
(\ref{Eq7}) in the region
\begin{equation}\label{Eq19}
r\gg C ,
\end{equation}
in which case one may approximate (\ref{Eq7}) by
\begin{equation}\label{Eq20}
\Big({{d^2}\over{dr^2}}+\omega^2-{{m^2-{1\over
4}}\over{r^2}}\Big)\psi_m=0\ .
\end{equation}
Here we have used the strong inequality 
$C^2m^2/r^4\ll m^2/r^2$ [see Eq. (\ref{Eq19})]. In addition, we have
used the strong inequality $Cm\omega/r^2\ll m^2/r^2$ which stems
from the small-frequency assumption $C\omega\ll 1\leq m$ [see Eq. (\ref{Eq10})].

The general mathematical solution of (\ref{Eq20}) can be expressed in terms of the Bessel
functions of the first and second kinds (see Eq. 9.1.49 of
\cite{Abram}):
\begin{equation}\label{Eq21}
\psi_m(r)=B_1r^{1\over 2}\cdot J_m(\omega r)+B_2r^{1\over 2}\cdot Y_{m}(\omega
r)\ ,
\end{equation}
where $\{B_1,B_2\}$ are normalization constants which, as we shall 
explicitly show below, can be determined by a functional matching procedure.
Using the small-argument,
\begin{equation}\label{Eq22}
\omega r\ll1\  ,
\end{equation}
asymptotic behaviors of the Bessel functions (see Eqs. 9.1.7 and 9.1.9 of \cite{Abram}), one finds from
(\ref{Eq21}) the expression
\begin{eqnarray}\label{Eq23}
\psi_m(r)=B_1(m!)^{-1}\Big({{\omega}\over{2}}\Big)^mr^{m+{1\over2}}
-B_2{\pi}^{-1}(m-1)!\Big({{\omega}\over{2}}\Big)^{-m}r^{-m+{1\over2}}\
\end{eqnarray}
for the radial acoustic eigenfunction that characterizes the linearized
perturbation modes of the hydrodynamic vortex model in the
intermediate radial region [see Eqs. (\ref{Eq19}) and (\ref{Eq22})]
\begin{equation}\label{Eq24}
C\ll r \ll 1/\omega\  .
\end{equation}

Interestingly, from Eqs. (\ref{Eq18}) and (\ref{Eq24}) one learns
that, for small resonant frequencies, there is an overlap radial
region which is determined by the strong inequalities
\begin{equation}\label{Eq25}
Cm\ll r_o\ll 1/\omega\  ,
\end{equation}
in which the expressions (\ref{Eq17}) and (\ref{Eq23}) for the
radial acoustic eigenfunction $\psi_m(r)$ of the hydrodynamic vortex model
are both valid. Note, in particular, that the two expressions
(\ref{Eq17}) and (\ref{Eq23}) for the eigenfunction $\psi_m(r)$
share the same radial functional behavior. One can therefore
determine the coefficients $\{B_1,B_2\}$ of the characteristic
radial eigenfunction (\ref{Eq21}) by matching the expressions
(\ref{Eq17}) and (\ref{Eq23}) in their overlap radial region (\ref{Eq25}). 
This functional matching procedure yields the expressions
\begin{equation}\label{Eq26}
B_1=-A_2{\pi}^{-1}(m-1)!m!\Big({{Cm\omega}\over{4}}\Big)^{-m}\ \ \
\end{equation}
and
\begin{equation}\label{Eq27}
B_2=A_1A_2B^{-1}_1
\end{equation}
for the normalization constants of the radial acoustic eigenfunction (\ref{Eq21}).

We are now in a position to derive the characteristic
resonance equation which determines the complex resonant frequencies
of the hydrodynamic vortex model. Using Eqs. 9.2.1 and 9.2.2 of
\cite{Abram}, one finds that the radial eigenfunction (\ref{Eq21})
is characterized by the large-$r$ asymptotic behavior
\begin{equation}\label{Eq28}
\psi(r\to\infty)=B_1\sqrt{2/\pi\omega}\cdot\cos(\omega
r-m\pi/2-\pi/4)+B_2\sqrt{2/\pi\omega}\cdot\sin(\omega
r-m\pi/2-\pi/4)\ .
\end{equation}
Taking cognizance of the boundary condition (\ref{Eq9}), which
characterizes the asymptotic spatial behavior of the radial
eigenfunctions of the hydrodynamic vortex model, one deduces from
(\ref{Eq28}) the simple relation
\begin{equation}\label{Eq29}
B_2=iB_1\  .
\end{equation}
Substituting (\ref{Eq29}) into (\ref{Eq27}), one finds the relation
\begin{equation}\label{Eq30}
iB^2_1=A_1A_2\  ,
\end{equation}
which yields the compact resonance equation [see Eq. (\ref{Eq26})]
\begin{equation}\label{Eq31}
\Big({{Cm\omega}\over{4}}\Big)^{2m}=i\Big[{{(m-1)!m!}\over{\pi}}\Big]^{2}\cdot{{A_2}\over{A_1}}
\end{equation}
for the complex resonant frequencies that characterize the dynamics
of linearized perturbation modes in the hydrodynamic vortex model.
It is worth emphasizing again that the analytically derived resonance condition
(\ref{Eq31}) is valid in the low frequency regime [see Eq.
(\ref{Eq25})] 
\begin{equation}\label{Eq32}
Cm\omega\ll1\  ,
\end{equation}
which corresponds to the small dimensionless ratio
\begin{equation}\label{Eq33}
{{A_2}\over{A_1}}\ll1\  .
\end{equation}
Since each inequality in (\ref{Eq25}) roughly corresponds to an order-of-magnitude
difference between two physical quantities [that is, $Cm/r_o\lesssim
10^{-1}$ and $r_o/(1/\omega)\lesssim 10^{-1}]$, the analytically
derived resonance condition (\ref{Eq31}) for the characteristic
resonant frequencies of the hydrodynamic vortex model is expected to
be valid in the dimensionless low frequency regime $Cm\omega\lesssim 10^{-2}$.

\section{The superradiant instability spectrum of the hydrodynamic vortex model}

As emphasized above, in the present analytical study we focus on the
low-frequency resonance spectrum which characterizes the dynamics of 
sound waves in the hydrodynamic vortex model. In particular, in the dimensionless regime
\begin{equation}\label{Eq34}
\omega R_0\ll m Z_{\omega}/\rho
\end{equation}
of small resonant frequencies, one can approximate the 
dimensionless ratio (\ref{Eq15}) by (here we have used Eq. 9.1.27c of \cite{Abram})
\begin{equation}\label{Eq35}
{{A_2}\over{A_1}}=-{{J^{'}_m\big({{Cm}\over{R_0}}\big)}\over{Y^{'}_m
\big({{Cm}\over{R_0}}\big)}}\  .
\end{equation}
It is worth noting that, in general, the frequency-dependent impedance diverges
as an inverse power law of the frequency in the small frequency
$\omega\to0$ limit \cite{Lax}. Thus, one finds that the ratio
$\omega R_0/(m Z_{\omega}/\rho)\to 0$ approaches zero faster than
$\omega^1$ in the low frequency regime that we explore analytically here \cite{Lax}. 
Note that the small-frequency relation (\ref{Eq34}) corresponds to the
Neumann-type boundary condition ${{d\Psi}\over{dr}}(r=R_0)=0$
\cite{Cars} for the linearized perturbation fields at the surface
$r=R_0$ of the central cylinder [see Eq. (\ref{Eq8})].

Substituting the dimensionless ratio (\ref{Eq35}) into the
analytically derived resonance equation (\ref{Eq31}), one finds the simple resonance relation
\begin{equation}\label{Eq36}
C\omega(C,R_0,m)={{4}\over{m}}\Big[{{(m-1)!m!}\over{\pi}}\Big]^{{1}/{m}}\cdot\Bigg|{{J^{'}_m\big({{Cm}\over{R_0}}\big)}
\over{Y^{'}_m\big({{Cm}\over{R_0}}\big)}}\Bigg|^{{1}/{2m}}\times
e^{\pm i\pi/4m}\  ,
\end{equation}
which characterizes the dynamics of linearized wave fields in
the effective acoustic spacetime (\ref{Eq3}). 
The $+/-$ signs in the analytically derived functional expression (\ref{Eq36}) refer respectively to negative/positive values of the dimensionless 
ratio $J^{'}_m(Cm/R_0)/Y^{'}_m(Cm/R_0)$. 
Here we have used the relation $\pm i=e^{i\pi(\pm{{1}\over{2}}+2n)}$,
where the integer $n$ is the resonance parameter of the acoustic
field mode.

Since the low frequency resonances of the hydrodynamic vortex model
that we explore in the present paper correspond to the strong inequality
(\ref{Eq33}), one deduces from (\ref{Eq35}) that our analytical
study is valid for central supporting cylinders whose radii lie in the vicinity
of the discrete critical radii \cite{Hodmarg}
\begin{equation}\label{Eq37}
R^*_0(C,m;k)={{Cm}\over{j^{'}_{m,k}}}\ \ \ \ ; \ \ \ \ k=1,2,3,...\
\end{equation}
for which $J^{'}_m(Cm/R^*_0)=0$ [and thus also 
$\omega[C,R^*_0(C,m;k),m]=0$, see Eq. (\ref{Eq36})], where
$j^{'}_{m,k}$ is the $k$th positive zero of the function $J^{'}_m(x)$ \cite{Abram,Bes}. 
As explicitly shown in \cite{Hodmarg}, the discrete set (\ref{Eq37}) of critical cylinder radii
support the marginally-stable static resonances (with
$\Re\omega=\Im\omega=0$) of the hydrodynamic vortex model. 

In particular, as shown numerically in \cite{Cars} and analytically in
\cite{Hodmarg}, the hydrodynamic vortex model is characterized by
the existence of a discrete set of critical cylinder radii,
$\{R^*_0(C,m;k)\}_{k=1}^{k=\infty}$, that support spatially regular
{\it static} ($\Re\omega=\Im\omega=0$) acoustic field configurations.
Interestingly, it has been shown \cite{Cars,Hodmarg} that, for given values of the fluid-field parameters
$\{C,m\}$, these marginally-stable field configurations mark the
onset of the exponentially growing superradiant instabilities in the
hydrodynamic vortex model.

\section{Summary and discussion}

The superradiant instability properties of the hydrodynamic vortex
model, an asymptotically flat horizonless acoustic geometry that 
possesses an ergoregion, were studied {\it analytically}. 
In particular, we have derived the compact analytical formula (\ref{Eq36}) for the parameter-dependent complex 
resonances of the composed fluid-cylinder system. 

Our analytical matching procedure is valid in the dimensionless small-frequency regime (\ref{Eq10}) and 
it is therefore convenient to define the dimensionless physical quantity
\begin{equation}\label{Eq38}
\Delta R\equiv {{R_0-R^*_0(C,m;k)}\over{R^*_0(C,m;k)}}\ \ \ \
\text{with}\ \ \ \ \Delta R\ll1\  ,
\end{equation}
in terms of which one can expand the small dimensionless ratio in Eq. (\ref{Eq35}) in the
form
\begin{equation}\label{Eq39}
{{J^{'}_m\big({{Cm}\over{R_0}}\big)}\over{Y^{'}_m\big({{Cm}\over{R_0}}\big)}}=
-j^{'}_{m,k}\cdot{{J^{''}_{m}(j^{'}_{m,k})}\over{Y^{'}_m(j^{'}_{m,k})}}
\cdot \Delta R\cdot[1+O(\Delta R)]\  .
\end{equation}
Here we have used the Taylor expansions
$J^{'}_m(Cm/R_0)=J^{'}_m(Cm/R^*_0)+J^{''}_m(Cm/R^*_0)\cdot(-Cm/R^*_0)\cdot\Delta
R+O[(\Delta R)^2]$ with $J^{'}_m(Cm/R^*_0)\equiv 0$ and
$Y^{'}_m(Cm/R_0)=Y^{'}_m(Cm/R^*_0)\cdot[1+O(\Delta R)]$. 
Using Eq. 9.1.31 of \cite{Abram}, one can write
${{J^{'}_m({{Cm}/{R_0}})}\over{Y^{'}_m({{Cm}/{R_0}})}}=
-j^{'}_{m,k}\cdot{{J_{m-2}(j^{'}_{m,k})-2J_{m}(j^{'}_{m,k})+J_{m+2}(j^{'}_{m,k})}
\over{2[Y_{m-1}(j^{'}_{m,k})-Y_{m+1}(j^{'}_{m,k})]}} \cdot \Delta R\cdot[1+O(\Delta R)]$ for the ratio (\ref{Eq39}), 
which is now expressed in terms of the Bessel functions themselves.

Substituting the ratio (\ref{Eq39}) into (\ref{Eq36}), one obtains the functional relation
\begin{equation}\label{Eq40}
C\omega(C,R_0,m)={{4}\over{m}}\Big[{{(m-1)!m!}\over{\pi}}\Big]^{{1}/{m}}\cdot\Bigg|j^{'}_{m,k}
\cdot{{J^{''}_{m}(j^{'}_{m,k})}\over{Y^{'}_m(j^{'}_{m,k})}}\cdot\Delta R\Bigg|^{{1}/{2m}}\times e^{\pm i\pi/4m}\  
\end{equation}
for the low frequency resonances that characterize the dynamics of
linearized sound modes in the effective acoustic spacetime, 
where the $+/-$ signs in (\ref{Eq40}) refer respectively to negative/positive values of 
the dimensionless quantity $\Delta R$. 
It can be checked directly that the coefficient
$-j^{'}_{m,k}{{J^{''}_{m}(j^{'}_{m,k})}\over{Y^{'}_m(j^{'}_{m,k})}}$
in (\ref{Eq39}) is a positive definite expression, which implies that the ratio 
${{J^{'}_m({{Cm}/{R_0}})}\over{Y^{'}_m({{Cm}/{R_0}})}}$ is positive/negative for positive/negative 
values of the dimensionless quantity $\Delta R$ [see Eq. (\ref{Eq39})]. 

Intriguingly, from the analytically derived functional relation (\ref{Eq40}) one deduces 
the characteristic inequality
\begin{equation}\label{Eq41}
\Im\omega>0\ \ \ \ \text{for}\ \ \ \ \Delta R<0\  , 
\end{equation}
which describes exponentially growing supperradiant instability modes 
of the hydrodynamic vortex model [see Eq. (\ref{Eq6})]. 

Finally, it is interesting to point out that the low frequency resonance expression (\ref{Eq40}) of the hydrodynamic
vortex model can be further simplified in the following two asymptotic regimes:

(1) In the $m\gg1$ limit of large harmonic indices, one finds 
\begin{equation}\label{Eq42}
C\omega(m\gg 1,k)={{4m}\over{e^2}}\cdot|\Delta R|^{{1}/{2m}}\times
e^{\pm i\pi/4m}\  .
\end{equation}
Here we have used the relations $(m!)^{1/m}\to m/e$ and 
$|j^{'}_{m,k}\cdot{{J^{''}_{m}(j^{'}_{m,k})}/{Y^{'}_m(j^{'}_{m,k})}}|^{{1}/{2m}}\to1$ 
in the asymptotic $m\gg1$ regime (see Eqs. 9.5.16 and 9.5.20 of \cite{Abram}). 
It is important to note that, taking cognizance of (\ref{Eq10}), one finds that
the asymptotic $m\gg 1$ formula (\ref{Eq42}) for the complex resonances of the
hydrodynamic vortex model is valid in the $\Delta R\ll m^{-2m}$ regime.

(2) In the $k\gg m$ limit of small cylinder radii, one finds using Eq. 9.5.13 of \cite{Abram}
\begin{equation}\label{Eq43}
j^{'}_{m,k}=k\pi[1+O(m/k)]\ \ \ \ \text{for}\ \ \ \ k\gg m\
\end{equation}
[this yields $R^*_0(C,m;k\gg m)={{Cm}\over{k\pi}}\ll C$, see Eq. (\ref{Eq37})], 
which yields [see Eq. (\ref{Eq40})]
\begin{equation}\label{Eq44}
C\omega(m,k\gg m)={{4}\over{m}}[(m-1)!m!]^{{1}/{m}}\cdot\Big|{{k}\over{\pi}}\cdot\Delta R\Big|^{{1}/{2m}}
\times e^{\pm i\pi/4m}\  .
\end{equation}
Here we have used the relation ${{J^{''}_{m}(j^{'}_{m,k})}
/{Y^{'}_m(j^{'}_{m,k})}}\to -1$ for $k\gg m$ (see Eqs. 9.2.1 and 9.2.2 of \cite{Abram}). 
It is important to note that, taking cognizance of (\ref{Eq10}), one finds that
the asymptotic $k\gg m$ formula (\ref{Eq44}) for the complex resonances of the
hydrodynamic vortex model is valid in the $\Delta R\ll k^{-1}$ regime.



\bigskip
\noindent
{\bf ACKNOWLEDGMENTS}
\bigskip

This research is supported by the Carmel Science Foundation. I thank 
Yael Oren, Arbel M. Ongo, Ayelet B. Lata, and Alona B. Tea for stimulating discussions.



\begin{thebibliography}{99}
\bibitem{Noteun} It is worth emphasizing that, as originally shown
by Unruh \cite{Unr}, the wave equation that 
governs the dynamics of linearized sound waves in physical
fluid systems [see Eq. (\ref{Eq7}) below] is mathematically analogous to the familiar Klein-Gordon wave
equation that governs the dynamics of linearized massless scalar
fields in curved spacetimes.
\bibitem{Notecc} Note that, in these units, the proportionality constant $C$ [see Eq. (\ref{Eq2})]
has the dimensions of length.
\bibitem{Notecy} Note that in (\ref{Eq6}) the effective scalar field $\Psi$ is expanded in terms of
linearized acoustic wave modes with cylindrical symmetry.
\bibitem{Notemp} Note that the azimuthal periodicity of the angular eigenfunction $e^{im\phi}$ 
implies that the azimuthal harmonic index $|m|$, which characterizes the linearized acoustic field
mode, is an integer.
\bibitem{Noteinv} It is worth pointing out that the ordinary differential equation
(\ref{Eq7}), which determines the spatial behavior of the radial acoustic eigenfunctions, 
is invariant under the symmetry transformation $m\to
-m$ with $C\to -C$. One can therefore assume, without loss of
generality, that $m>0$ and $C>0$.
\end{thebibliography}
\end{document}